\documentstyle[11pt,paspconf]{article}

\input epsf.tex

\begin{document}

\title{The Building up and Evolution of Galactic Disks}

\author{Claudio Firmani, Vladimir Avila-Reese, and Xavier Hern\'{a}ndez\altaffilmark{1}}
\affil{Centro de Instrumentos, U.N.A.M., \\
Apdo. Postal 70-186, 04510 M\'exico D.F., Mexico }
\affil{Instituto de Astronom\'{\i}a, U.N.A.M., \\
Apdo. Postal 70-264, 04510 M\'exico D.F., Mexico}

\altaffiltext{1}{Presently at Cambridge University, UK}



\begin{abstract}
 The formation and evolution of disk galaxies in the cosmological context is studied. We consider the observable properties of disk galaxies and treat the disk formation and galactic evolutionary processes in a self-consistent fashion. We find the matter accretion regime to be the dominant ingredient in establishing the Hubble sequence. The accretion regime is a phenomenon directly related to the statistical properties of the primordial density fluctuations from which disk galaxies emerged.
\end{abstract}




\section{Introduction}

Disk galaxies exhibit well defined observational properties and correlations, many of  which are on the basis of the Hubble sequence ($HS$). The origin of such sequence is an outstanding problem of astrophysics. There are two directions in which this problem has been studied: considering the $HS$ as a result of primordial initial conditions or (and) of secular disk evolution. Since galaxies are
not only individual ''ecosystems'' where stars are born, live, and die, but also
the structural unities of the universe as a whole, the origin of at least
some of their physical properties should be related to the cosmological
initial conditions. Only through a combined study, where the evolutionary processes of the 
visible galactic system as well the cosmological initial conditions are taken into account, it will be possible to
distill the main ingredients that establish the $HS$. This is the line we follow in order to construct a scenario of galaxy formation and
evolution mainly focused to disk galaxies.

\subsection{Gas accretion in disk galaxies}

The relevance of gas accretion on the evolution of galactic disks has been pointed out repeatedly. Discrepancies between
observations and the simple ''closed-box'' evolutionary models like the
G-dwarf problem, the fast gas consumption, the gas depletion paradox, and
the systematic variation of the star formation ($SF$) time-scale along the $%
HS$, can be solved stating that disk did not form instantaneously but have
grown by a gradual process of accretion, which is fast at the initial epochs
for early type disks and slower for late-type disks.

Likewise, evolutionary models where the $SF$, hydrodynamics, and
gravitational interactions (including dark matter) of a galactic disk are
treated in a self-consistent fashion (Firmani $et$ $al.$ 1996a, $FHG,$ see
also Firmani \& Tutukov 1992, 1994) have shown that late-type disks can not
be obtained without gas accretion.

\subsection{Gravitational collapse of primordial fluctuations}
In the frame of the most predictive inflationary cosmological models, where baryon matter is only a small fraction of cold dark
matter, cosmic structure develops by a process of continuous merging and
accretion. Because galactic disks are cold and thin structures they could
not have suffered major mergers (T\'{o}th \&\ Ostriker 1991). On the other
hand disk galaxies are typically located in low density environments where
mayor mergers are rather rare. That is why we envisage disk galaxy
formation as the process of gravitational collapse of individual density
profiles around the local maxima of the fluctuation field; of course some
substructure (merging) will be present, but we treat it as a second-order
ingredient.

\vspace{0.3in}

 The possible mass growth histories of a given present-day galaxy can roughly
be estimated through a particular physical interpretation of the conditional
probabilities found by Bower 1991, Bond $et$ $al.$ 1991, and Lacey \& Cole 1993.
We calculate the virialization of dark halos following these histories or
accretion regimes assuming spherical symmetry (see Avila-Reese \& Firmani
1996, $AF$). Galactic disks are built up into the dark halos assuming detailed
conservation of angular momentum distribution. The evolution of accreting
disks is followed through a previously developed numerical code ($FHG$).
In this way, we explore how do the different cosmological accretion regimes
influence the evolution of galactic disks devoting special attention to disk
structure, $SF$ history, Hubble type, rotation curve and the Tully-Fisher
relation.

\section{Initial conditions and formation and evolution of galactic disks}

 We are interested in studying the gas mass and angular momentum accretion
regimes over the galactic disks. As was mentioned above, our main assumption
is that these regimes are related to the cosmological collapse of primordial
density fluctuations. For Gaussian random fields, it is possible to
statistically derive mass growth histories. We use the Monte Carlo formalism
as in Lacey \& Cole and for a given mass we found the average trajectory and
the two still statistically sifnificant average deviations from the most possible case ($AF$). Since the mass growth histories are fixed to the present-day mass,
these trajectories reflect the range and distribution of possible accretion
regimes of disk galaxies (see fig. 1 of $AF$). The regimes are characterized
by the parameter $\gamma (t)\equiv \frac{d\ln M(t)}{d\ln t}$ which is measured from
today back to the past. For every mass one has a low, average, and high
accretion regimes which slowly change with the mass ($AF$).

The virialized structures which arise according to these accretion regimes
are calculated using an extension of the secondary infall mechanism (see $AF$).
 The formation of disks in the dark halos implies gas energy dissipation and some initial angular momentum. We do not treat
the gas cooling and assume that as soon as a shell virializes, the baryon mass
fraction cools and falls to the center in a dynamical time. The disk is
built up with the infalling gas from the evolving dark halo. Concerning the angular momentum, it is widely accepted that galaxies acquire it in the linear regime through the  tidal torques due to the neighboring density fluctuations. Because in our approximation the halos are spherically symmetric (the moment of inertia vanishes) we can not calculate the value of the angular momentum. However, using the Zel'dovich approximation it is possible to calculate the dependence of the acquired angular momentum on the distance from the center of the collapsing region. A simple extension of a spetial formalism (e.g. see White 1994) was applied in order to obtain this relation, and the value of the tidal torque (a free parameter) was fixed through a normalization to the Galaxy.
It is encouraging that the spin parameter $\lambda $ predicted in numerical
simulations is in agreement with such a normalization.

\begin{figure}
\vspace{9.5cm}
\includegraphics{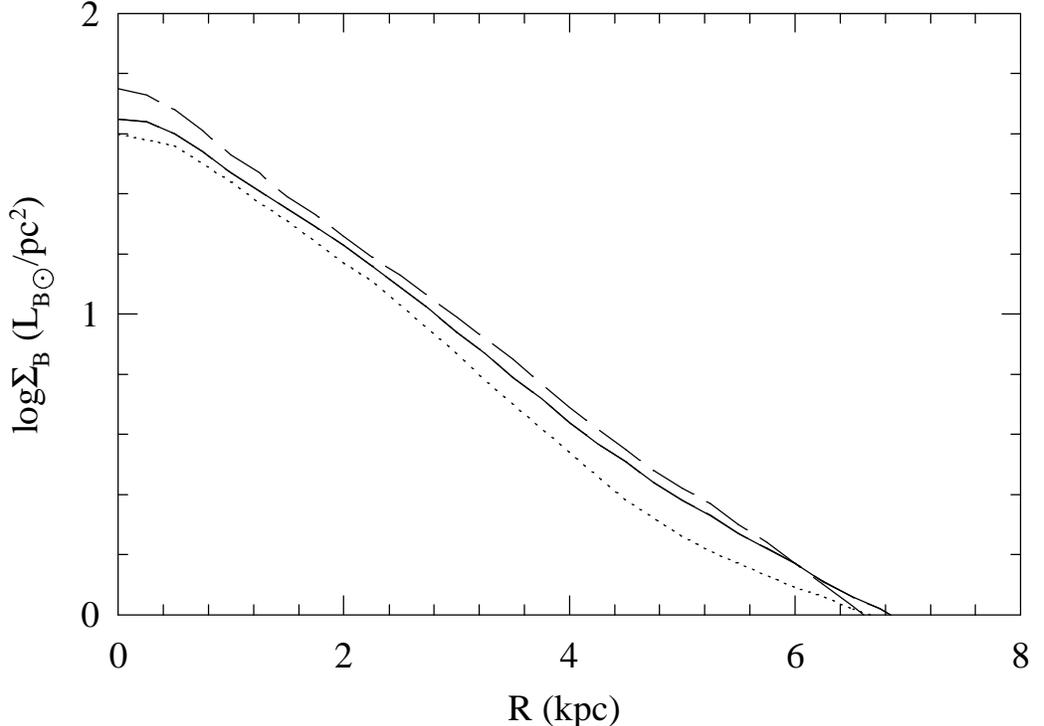}
\caption{The B radial surface brightness profiles of a $10^{11}M_{\odot }$ galaxy. Dashed, solid, and dotted lines correspond to the low, average and high accretion regimes, respectively} 
\end{figure}

Under the assumption of detailed angular momentum conservation, the gas falls just to the radius where its angular momentum provides centrifugal support, resulting in a disk mass distribution which tends to be exponential ( e.g. Gunn
1981, 1987). We tested the robustness of the result introducing several
departures from this approach where the total angular momentum was conserved
but some radial transfer of the falling gas was allowed. An exponential
disk was ever the result of this gradual collapse.

In the forming disks, galactic evolutionary processes will arise. $SF$,
energy injection to the $ISM$, turbulent dissipation, self-gravity and
gravitational instabilities, and their non-linear interactions are some of
the crucial ingredients which regulate the evolution of galactic disks. A
theoretical approach where $SF$ is driven by gravitational instabilities and
regulated by energy balance between $SN$ input and turbulent dissipation is
applied ($FHG$). Gravity produced by the forming disk and dark halo and the
corresponding interaction between them is treated in detail. In the cases of
slowly growing disks, after some short time-scales, the $SF$ rate becomes
proportional to the gas accretion rate, while disks formed by a fast
collapse, where the gas is supplied only by the mass loss from stars, will have a declining $SF$ rate with time

\begin{figure}
\vspace{9.0cm}
\includegraphics{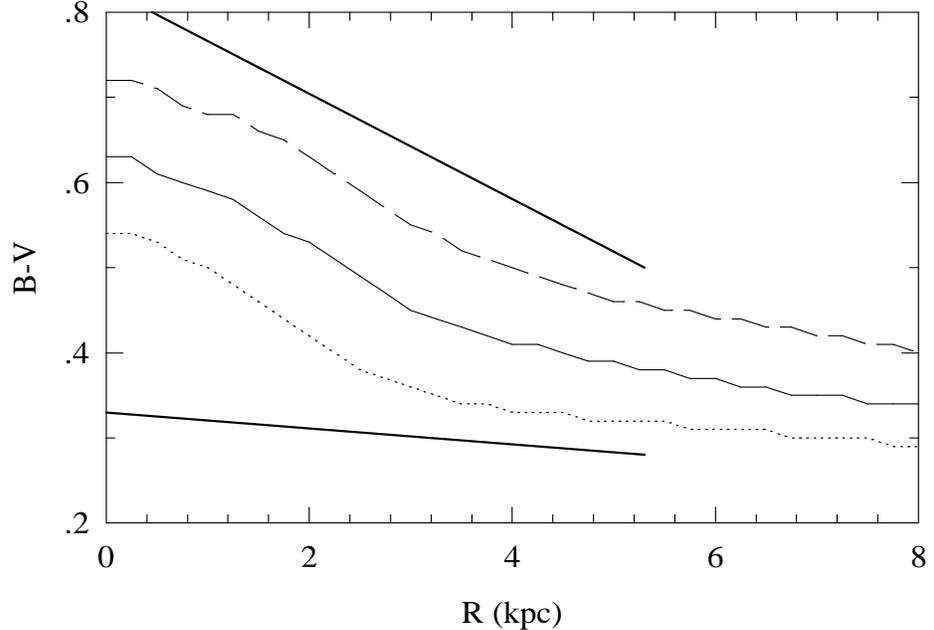}
\caption{Radial B-V color profiles for a 10$^{11}M_{\odot}$ galaxy.The line simbols are the same as in fig.1. For explanations regarding the two extreme  solid lines see the text}
\end{figure}

\begin{figure}[t]
\vspace{9.0cm}
\includegraphics{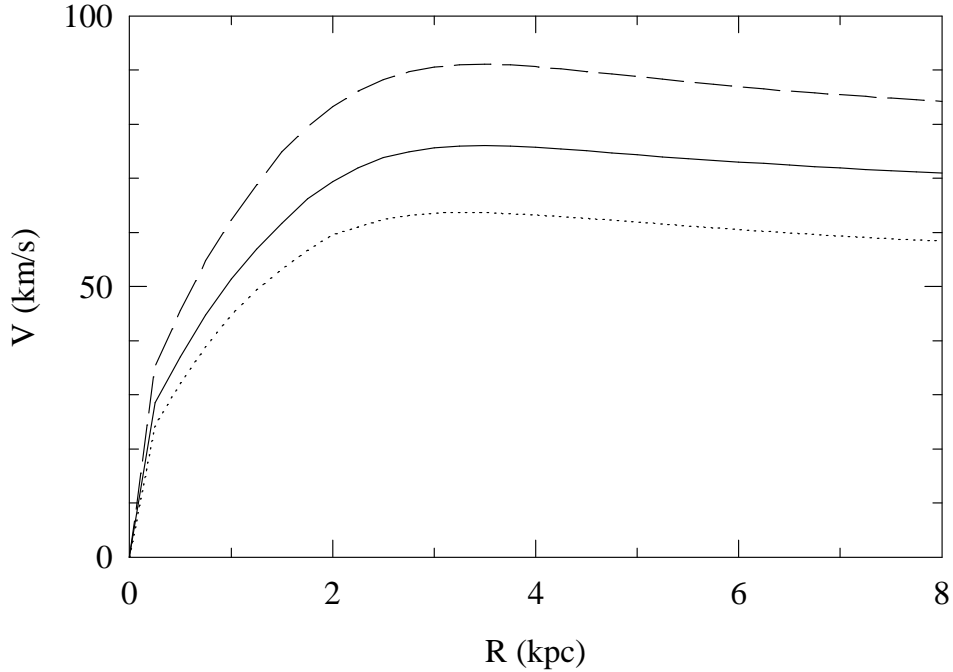}
\caption{Rotation curves for a 10$^{11}M_{\odot }$ galaxy. The line simbols are the same as in figure 1}
\end{figure}

\begin{figure}
\vspace{9.0cm}
\includegraphics{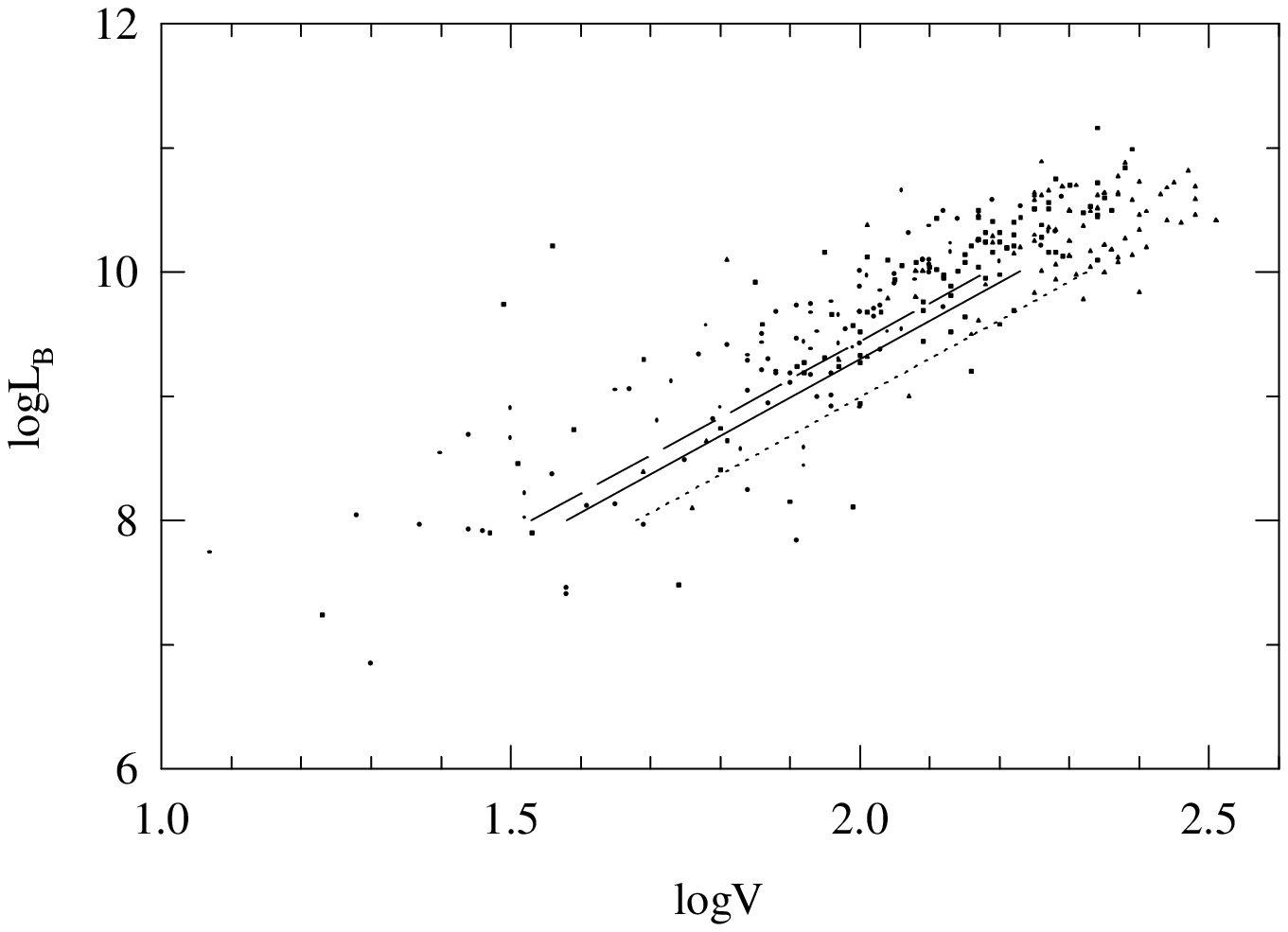}
\caption{The Tully-Fisher relation in B-luminosty for a sequence of models corresponding to the high (dashed line), average (solid line), and low (dotted line) accretion regimes. The points are the observational data estimated from a compilation of the RC3 and Tully catalogs (see Firmani \& Tutukov 1994). The velocity is not that which was obtained from direct observations, and the points only approximately reflect the Tully-Fisher relation}
\end{figure}

\section{Results}

Using the approach described above and in $AF$ we present here some
preliminary results on disk formation and evolution in the cosmological
context where $a$ $priori$ some key ingredients related to the initial
conditions, the evolutionary processes, and the observational properties ($%
HS $) where remarked and self-consistently combined in order to construct a
scenario which we hope will help us to better understand the real phenomenon.

Our calculations are made here for a standard cold dark matter model, and
the results are shown for a $10^{11}M_{\odot }$ galaxy.

Figure 1 shows the exponential behavior of the surface brightness for the
three average mass growth histories. Concerning the $SF$ history the models
predict gentle radial gradients in the $B-V$ colors (figure 2) in agreement
with recent observations (de Jong 1995). The average colors of the disk
change with the accretion regime going from $B-V\approx 0.4$ for high
accretion regimes to $B-V\approx 0.7$ for low accretion regimes. The angular
momentum transfer from the infalling gas to the dark halo may influence the
blue limit, while the red limit is fixed basically by our library of red
giant evolutionary tracks. The extreme cases of a sudden initial $SF$ burst and a constant $SF$ history on the Hubble time are shown in Figure 2 with thick solid lines.

In Figure 3 the final rotation curves are presented. They retain some
information about the primordial fluctuations notwithstanding the fact that the
contraction of baryon matter has altered the gravitational potential in the
visible galactic regions. It is clearly seen that galaxies formed by a fast
collapse are more concentrated than those emerged from an extended collapse.

The gravitational contribution of dark matter in the center appears to be
sensible to the initial kinetic energy content (see also $AF$). In order to
be in agreement with published decompositions of rotation curves we have to
fix a high kinetic energy content.

The results of our models are in agreement with the Tully-Fisher relation
(figure 4) providing evidence about its cosmological origin.

\section{Conclusions}

In the scenario for disk galaxy formation and evolution proposed here,
galaxies arise from the gravitational collapse of density profiles around
local maxima of the primordial density field in an expanding universe. The role of substructure (merging) is considered as a second-order phenomenon. The
initial conditions generated in agreement with this vision were used in
galactic evolutionary models which finally predicted present-day galactic
disks with exponential blue luminosity profiles, color radial gradients,
near flat rotation curves, and a Tully-Fisher relation in blue with a slope
of $\sim 3.3$. The calculated properties that can define the $HS$ are the
integral $B-V$ color, the $SF$ history, the present gas content, and the
degree of compactness of the galaxies. All these properties have shown to
follow a sequence along the range of cosmological accretion regimes
suggesting that this could be the dominant ingredient in establishing the $HS
$. Although we have not yet made detailed calculations it is easy to identife the control of the accretion regime also on the old stellar spheroid
component which can be formed at the beginning of the evolution by a violent
collapse (Firmani $et$ $al.$ 1996b).

\acknowledgments
We thank Antonio Garc\'es for technical help in preparing some of the figures, and Gabriella Piccineli for a critical reading of the manuscript. V.A-R. received partial finantial support from the PADEP-UNAM program. He akcnowledges a fellowship from CONACyT under the iberoamerican program MUTIS.

\end{document}